# Coulomb-blockade-controlled single-electron point source

Victor I. Kleshch[1,*], Vitali Porshyn[2], Anton S. Orekhov[3], Andrey S. Orekhov[3,4],

Dirk Lützenkirchen-Hecht[2], Alexander N. Obraztsov[1,5]

[1]*Department of Physics, Lomonosov Moscow State University, Moscow 119991, Russia*

[2]*Physics Department, Faculty of Mathematics and Natural Sciences, University of Wuppertal, Wuppertal 42119, Germany*

[3]*Shubnikov Institute of Crystallography of FSRC "Crystallography and Photonics" of the Russian Academy of Sciences, Moscow 119333, Russia*

[4]*Electron Microscopy for Materials Science (EMAT), Department of Physics, University of Antwerp, Antwerp B-2020, Belgium*

[5]*Department of Physics and Mathematics, University of Eastern Finland, Joensuu 80101, Finland*

* Corresponding author, e-mail: klesch@polly.phys.msu.ru

**Coulomb blockade is a fundamental phenomenon in physics enabling transfer of individual electrons one by one into electrically isolated nanostructures such as nanowires or quantum dots[1,2] and thereby creation of sources of single electrons[3,4]. Nowadays, solid-state single-electron sources are key elements of the emerging new technologies of quantum information processing and single-electron electronics[5]. Moreover, advanced research in free-electron quantum optics[6] and developments in electron microscopy[7] require the point sources of free electrons in vacuum[8,9], which can be controlled on a single electron level[10]. However, up to now, single-electron vacuum guns were not realized in practice. The problems to be solved include the formation of stable tip-shaped heterostructured emitters[11,12] and control of liberated electrons in time and energy domain[13]. Here we overcome these challenges by creating a field emission (FE) electron source based on a carbon nanowire coupled to an ultra-sharp diamond tip by a tunnel junction. Using energy spectroscopy, we directly observe**

**Coulomb oscillations of the electron Fermi level in the nanowire at room temperature and FE currents up to 1 μA. We reveal that the oscillations are suppressed at high FE current either by Joule heating or due to the high tunneling resistance, depending on the nanowire charging time, ranging from about 25 fs to 1 ps. We anticipate that the combination of introduced carbon single-electron sources with laser-induced gating[14] is highly promising for the creation of coherent ultrashort free-electron bunches of interest for low-energy electron holography[15] and ultrafast electron or X-ray imaging and spectroscopy[16,17].**

A typical FE electron source (cathode) under study is shown in Fig. 1a. It was fabricated by a FE-assisted structural modification of a micrometer-scale needle-like diamond crystal mounted on a tungsten tip (see Methods for details). It consists of a carbon nanowire extending from an amorphous carbon (a-C) layer which covers the diamond surface. The FE properties of the cathodes were investigated in an ultrahigh vacuum (UHV) setup, in which strong field at the nanowire apex was created by a positively biased mesh gate (Fig. 1b). The electrons emitted from the nanowire were transmitted through the gate and collected by a hemispherical analyzer. The total current, $I$, and the total electron energy distribution, $J(\varepsilon)$, were measured as a function of the gate voltage, $V$. Figure 2a shows a typical current-voltage, $I(V)$, dependence for one of the investigated carbon nanowire emitters. The most remarkable feature is a clear staircase-like pattern, which was observed in the $I(V)$ curves plotted in semi-logarithmic coordinates. The variations in FE are more evident from the dependence of the normalized differential conductance, $(dI/dV)/(I/V)$, on the applied voltage, demonstrating oscillations with a constant period, $\Delta V$. Similar staircase $I(V)$ behavior was reproducibly obtained for more than 10 samples with $\Delta V$ varying from 8 to 250 V, depending on the fabrication conditions. Moreover, by means of a high-current processing, it was possible to gradually shorten (or lengthen) the carbon nanowire and, correspondingly, increase (or decrease) $\Delta V$ in a certain range (see Supplementary Section I). An example of such modification is presented in Fig. 2b, which shows the results of three subsequent measurements demonstrating $\Delta V$ decrease with an increase in the nanowire length, $L$.

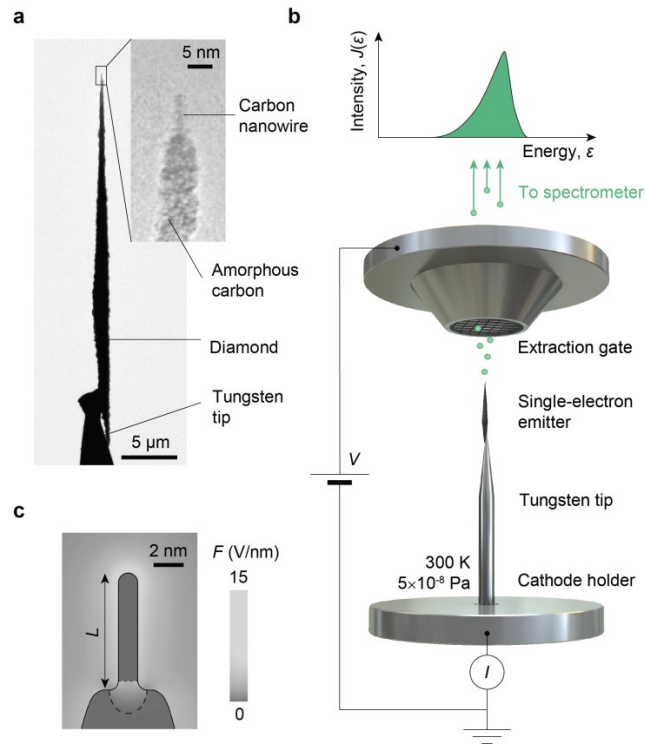

**Fig. 1. Heterostructured carbon field emitter and experimental scheme. a,** Transmission electron microscopy (TEM) images of a field emitter consisting of a carbon nanowire attached to a diamond needle-like crystal covered by an a-C layer. **b,** Schematic diagram of the experimental setup. Electrons are field emitted one by one from a single-electron source under a DC voltage, $V$, applied to the mesh gate electrode. The total energy distribution of the emitted electrons, $J(\varepsilon)$, is measured using a hemispherical spectrometer. The FE current, $I$, is measured by a picoammeter connected to the cathode. **c,** Simulated distribution of the electric field strength, $F$, for the emitter shown in (a) at a gate voltage $V = 80$ V (see Supplementary Section V). $L$ is the length of the nanowire. The dashed lines mark the boundaries of a depletion zone, formed at the junction of the nanowire and the a-C layer. Single-electron effects in FE from the nanowire are possible due to the formation of a double-barrier structure with a Schottky-type barrier in the depletion zone and a FE-barrier at the nanowire/vacuum interface.

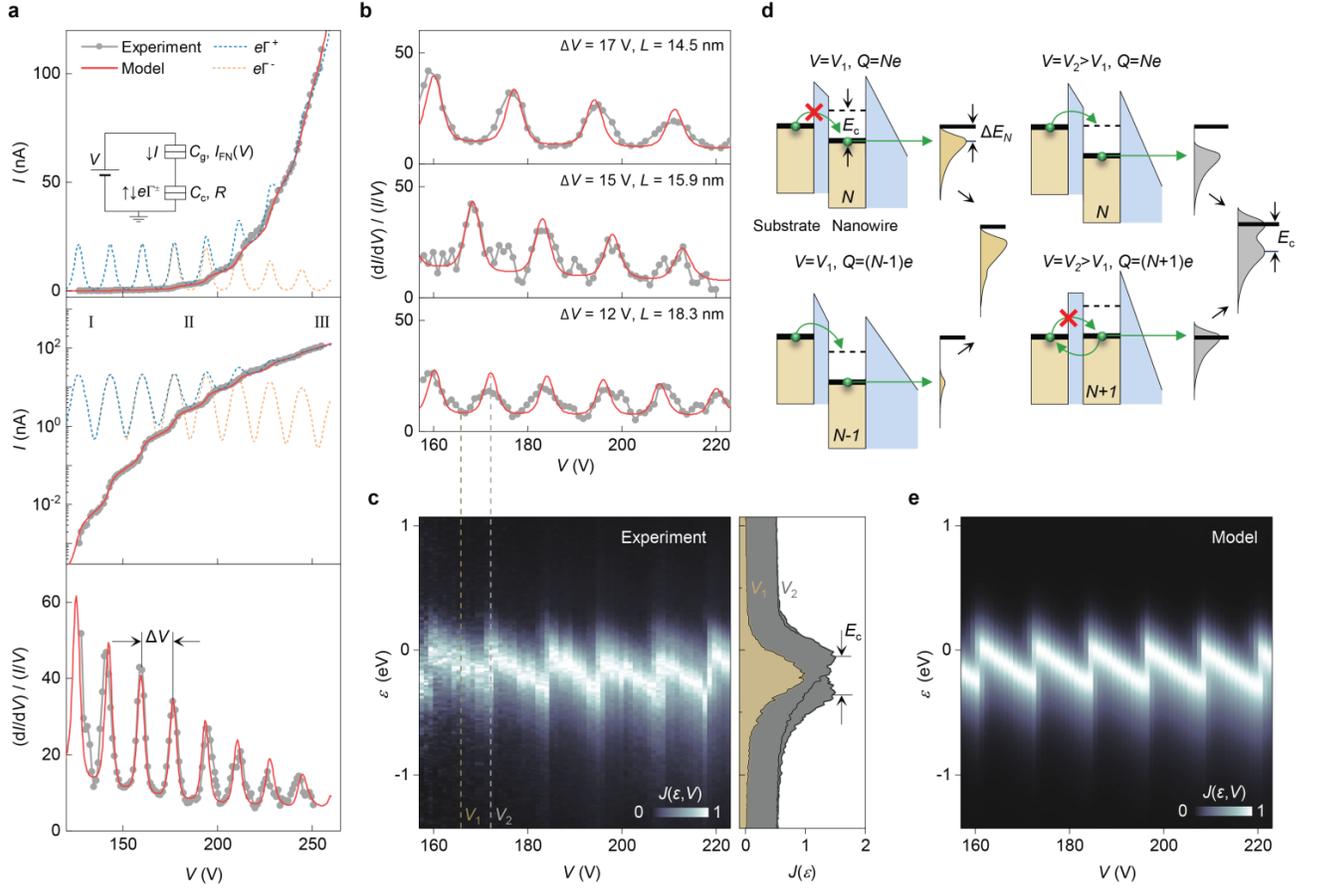

**Fig. 2. Coulomb oscillations in FE from a carbon nanowire. a,** Experimental and simulated staircase-like current-voltage characteristics, $I(V)$, in linear (upper panel) and semi-logarithmic (middle panel) coordinates, and the corresponding oscillations of the normalized differential conductivity (lower panel), $(dI/dV)/(I/V)$, with a period of $\Delta V$. The dashed lines are the dependences of $e\Gamma^{\pm}$ on $V$, where $e$ is the elementary charge and $\Gamma^{\pm}$ are the simulated average tunneling rates in the junction between the nanowire and the substrate. The inset shows an equivalent circuit of a double-barrier system consisting of a tunnel barrier between the nanowire and the substrate with resistance $R$ and a FE-barrier, which exhibits a Fowler-Nordheim-type (FN-type) current-voltage characteristic, $I_{FN}(V)$. $C_c$ and $C_g$ are the capacitances of the nanowire with respect to the rest of the cathode and to the gate, respectively. **b,** Three subsequent measurements (gray circles) of the differential conductivity with different periods, $\Delta V$, and their fits (red curves). The decrease in $\Delta V$ is associated with the increase in the nanowire length, $L$, due to the growth process, which was performed between the measurements. **c,** Sawtooth-like oscillations in the normalized total energy distribution map, $J(\varepsilon, V)$, measured together with the differential conductivity presented in the lower panel of (b). The dashed lines show voltages $V_1$ and $V_2$ at which the cross-sections (right panel) are made. **d,** Energy diagrams of the double-barrier system at two voltages $V_{1,2}$ defined in (c) and different charges, $Q$, of the nanowire (see main text). **e,** Simulated total energy distribution map corresponding to the experimental data in (c).

The staircase-like increase of FE current was previously observed in particular conditions for carbon nanotubes[12,18,19], while this is not typical for FE phenomenon in general. However, the most striking result obtained in our experiments is the dependence of electron energy spectra on the gate voltage. Figure 2c presents the normalized total energy distributions visualized in a two-dimensional map, $J(\varepsilon, V)$, where $\varepsilon$ is the kinetic electron energy relative to the Fermi level of the cathode electrode (see Methods). In the region between two maxima of the differential conductance, the spectrum consists of a single peak, as demonstrated by the map cross-section at $V_1$. The peak shifts downward in energy with increasing voltage until the maximum of the differential conductance is reached at $V_2$. Here, another peak appears at $\varepsilon=0$ eV and continues to shift with increasing voltage, while the lower-energy peak disappears. Thus, the peak position oscillates with the gate voltage and follows a remarkable "sawtooth-like" dependence with a period $\Delta V$.

The double-peak spectrum at $V_2$ indicates the presence of two different energy states, which could be associated with quantum confinement. However, confined field emitters, e.g. adsorbed atoms[20] or nanoscale metal protrusions[21], usually exhibit a continuous shift of discrete energy peaks with the applied voltage (because of the field penetration effect), rather than periodic oscillations. The principal novelty of our FE structure is that the nanowire is electrically isolated from the a-C layer by a potential barrier (Fig. 1c) formed due to the difference in concentrations of $sp^2$-bonded carbon atoms (Supplementary Section III). As a result, a double-barrier structure arises, in which Coulomb blockade effect becomes possible, since the addition of an extra electron to the nanowire, having a very low capacitance, requires a considerable energy $E_c$ (charging energy). Thus, in this case, FE is affected by the quantization of charge, rather than the size quantization, which is negligible compared to the Coulomb interaction in our carbon nanowires.

The energy diagram explaining transport in the double-barrier structure is presented in Fig. 2d. The nanowire is isolated by a tunnel barrier from the substrate, which has a fixed Fermi level. At voltage $V_1$ the nanowire appears mainly in two states with $N$ and $N$-1 electrons. In the $N$- electron state the transfer of an electron from the substrate to the nanowire is energetically unfavorable, since with the addition of an electron the nanowire Fermi level increases by $E_c$ and, thus, exceeds the substrate Fermi level. Therefore, the nanowire remains in the *N*-electron state until an electron tunnels into vacuum. In the (*N*-1)-electron state, electron tunneling into vacuum is much less probable than the substrate-nanowire tunneling. As a result, at $V=V_1$, FE predominantly occurs from the *N*-electron state and the spectrum consists of a single peak shifted relative to the substrate Fermi level by $\Delta E_N$, which is the Coulomb energy difference for the nanowire with *N* and *N*-1 electrons. The nanowire Fermi level defined by $\Delta E_N$ shifts downward with the voltage increase. At $V=V_2$ it aligns with the cathode Fermi level ($\Delta E_N=E_c$) and the transition to the (*N*+1)-electron state becomes favorable. In this case, FE occurs from both states with *N* and *N*+1 electrons and, correspondingly, two peaks separated by $E_c$ are observed in the energy spectrum.

Obtained experimental data are well reproduced by the model of the double-barrier system with an equivalent circuit presented in the inset of Fig. 2a (see Methods). Figure 2 shows the simulated spectra and fits of *I*(*V*) curves, from which we obtained the model parameters, including the dimensions of the nanowire, substrate-nanowire junction resistance, *R*, and the nanowire total capacitance, *C*. It should be noted that the curves in Fig. 2b, are perfectly fitted using the same set of parameters with the only variable being the nanowire length, *L*. This is clear evidence that the changes in the oscillation period $\Delta V$ in the experiment can be associated with the variation of *L*. It is also important to note that the fitting results coincide with the results of electrostatic simulations of a nanowire, based on its dimensions estimated from TEM images. In particular, the simulated nanowire-gate capacitance, $C_g$, obtained from the electric field distribution (Fig. 1c), coincides well with the analytical formula $C_g=e/\Delta V$ given by the model (see Supplementary Section V). This good agreement validates the assumption that the nanowire is isolated from the substrate.

To elucidate the mechanism of Coulomb oscillations, we calculated the average tunneling rates between the substrate and nanowire, $\Gamma^{\pm}$, which are related to the FE current as $I=e(\Gamma^+-\Gamma^-)$ (see

Methods). $\Gamma^{\pm}(V)$ dependences are shown in Fig. 2a, where three voltage regions can be distinguished. At low voltages (Region I), both $e\Gamma^{\pm}$ significantly exceed the FE current. At intermediate voltages (Region II), the FE barrier becomes more transparent and the substrate–nanowire tunneling rate $\Gamma^{+}$ increases with $I$, while the reverse tunneling rate $\Gamma^{-}$ remains approximately unchanged. At high voltages (Region III), Coulomb oscillations are gradually suppressed and $I \approx e\Gamma^{+}$, while $\Gamma^{-}$ decreases. A similar behavior of $\Gamma^{\pm}(V)$ was observed for all samples, however, the absolute values of $\Gamma^{\pm}$ varied significantly. $\Gamma^{\pm}$ can be estimated roughly by the reciprocal of the charging time constant $\Gamma^{\pm} \sim (RC)^{-1}$. For different samples we obtained $RC$ in the range from 25 fs to 1.5 ps. In the following, we consider two representative cases of the highest and lowest $RC$ values, shown in Fig. 3 and Fig. 4, respectively.

Figure 3 presents the results for the sample with a large $RC$ of 1.5 ps, which results from a high $R$ value of 3 MΩ. In this case an overall shift of the spectrum with voltage (Fig. 3a) was observed, as a result of an additional voltage drop, $IR_0$, inside the poorly conductive diamond needle with a large resistance of $R_0 \sim 90$ MΩ. More importantly, the energy peaks which correspond to the states with different numbers of electrons were found to coexist in a certain voltage region, see e.g. $N+4$ and $N+5$ states in Fig. 3b. Here, the tunneling rates through both barriers become comparable, i.e. $RC \sim e/I$, where $e/I$ is the average time between subsequent FE events. The number of coexisting electron levels increases with FE current, and the Coulomb oscillations become completely suppressed for $RC \gg e/I$. This is demonstrated in Fig. 3c, where we also present two other characteristic time intervals: the uncertainty time, $\tau_q$, and tunneling time, $\tau_t$. The Coulomb-blockade theory[2] claims that $RC \gg \tau_q \gg \tau_t$, is a necessary condition for the observation of single-electron charging effects. The above analysis shows that for the system with a FE barrier an additional condition $e/I \gg RC$ must hold as well.

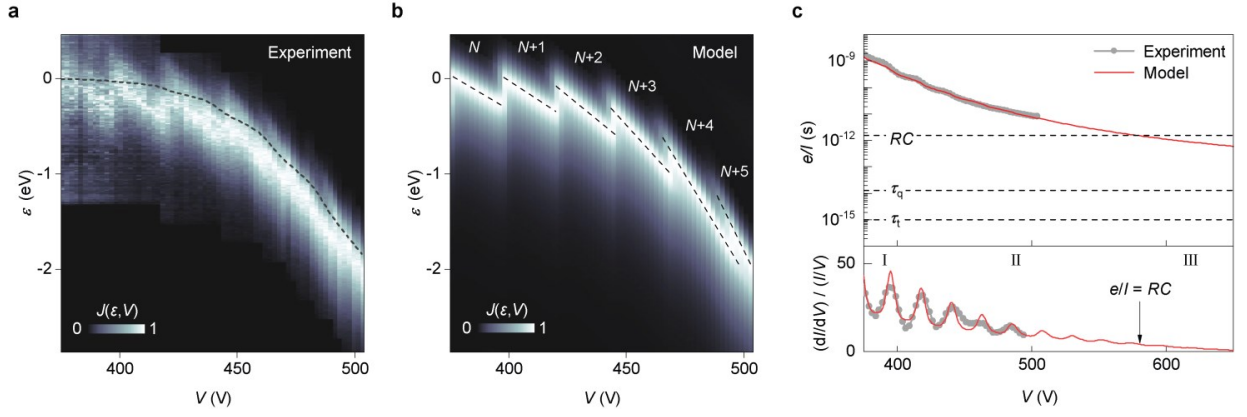

**Fig. 3. Coulomb oscillations for a large-*RC* field emitter. a,** Experimental total energy distribution map. The dashed line is the electron energy loss, $eIR_0$, associated with the voltage drop inside the diamond needle with a resistance $R_0$. **b,** Simulated total energy distribution map obtained at $RC = 1.5$ ps. The dashed lines are the positions of energy peaks corresponding to the states with a different number of electrons. **c,** The average time between electron FE events, $e/I(V)$, and the corresponding normalized differential conductivity, $(dI/dV)/(I/V)$. The fits (red lines) are obtained using the same parameters as in (b). The dashed lines represent the *RC*-time constant, the uncertainty time $\tau_q$ and the tunneling time $\tau_t$. The uncertainty time equals $\tau_q = R_q C$, where $R_q = e^2/h \sim 25.8$ kOhm is the quantum unit of resistance. It is associated with the time-energy uncertainty relation $E_c \tau_q \sim h$, with a charging energy $E_c = e^2/C$. The tunneling time, $\tau_t$, is the time spent by tunneling electron under the barrier. It is generally assumed to be of about $10^{-15}$ s.

The most important feature for emitters with small *RC* values is the strong broadening of the spectra with the applied voltage, as demonstrated in Fig. 4a for a sample with $RC=25$ fs. Performed modeling shows that the broadening is associated with an increase in the emitter temperature, *T*, due to Joule heating (Supplementary Section VI). The $T(V)$ dependence was directly extracted from the fits of the spectra and it was found that the oscillations almost completely disappear at $T>T_C=850$ K (Fig. 4c). Thus, unlike the large-*RC* case, here the single-electron effects are suppressed due to an increase in the thermal fluctuation energy, $k_B T$, which becomes comparable with the charging energy $E_c$, namely $E_c \sim 10 k_B T_C$, in agreement with the Coulomb-blockade theory[2]. The critical temperature $T_C$ is reached at *I* of about 1 μA and, correspondingly, $e/I$ is of about 0.2 ps, indicating that the FE-barrier effective tunneling rate is of about 5 THz. It is also worth noting that oscillations period here reaches 250 V. These remarkable characteristics, that are difficult to observe by employing conventional Coulomb-blockade-controlled solid-state devices, are achieved here due to the special geometry of the system (the drain is at a macroscopic distance from the source) and a small size of the nanowire with the high charging energy and short $RC$ time.

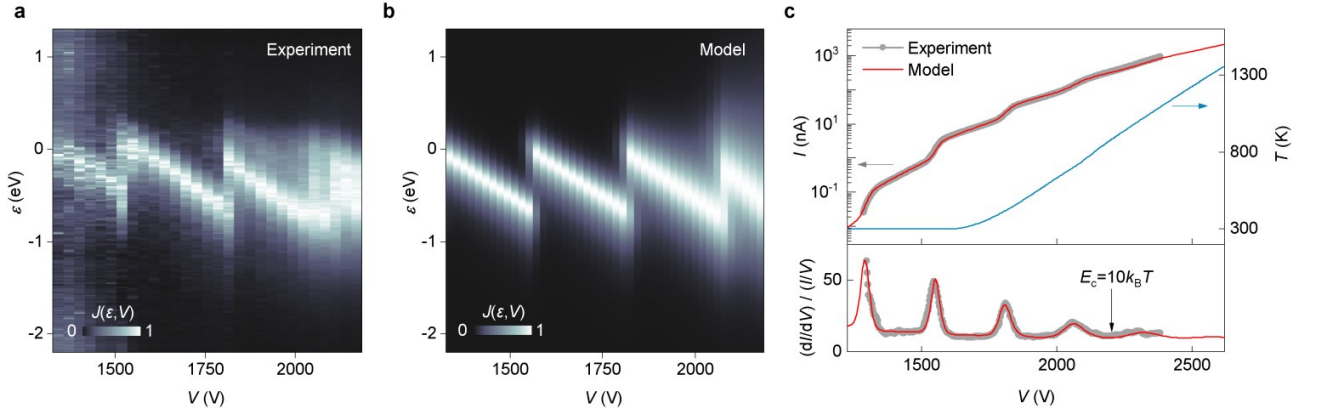

**Fig. 4. Coulomb oscillations for a small-$RC$ field emitter. a,** Experimental total energy distribution map, showing the broadening of the spectra with increasing voltage. **b,** Simulated total energy distribution map obtained at $RC = 0.025$ ps using $T(V)$ dependence. **c,** Current-voltage curve, $I(V)$, and the corresponding normalized differential conductivity, $(dI/dV) / (I/V)$. The blue curve is the emitter temperature dependence, $T(V)$, obtained by fitting the experimental spectra (See Supplementary Section VI). The fits (red lines) are obtained using the same parameters as in (b).

In conclusion, we report on the Coulomb blockade in an all-carbon heterostructured field emitter. Due to the strong carbon–carbon bonding, stable Coulomb oscillations persist at large currents, high tunneling frequencies and above room temperature. The small emission area and one-by-one electron tunneling make the liberated electrons more coherent both in space[22] and in time[23] in comparison with conventional metal-tip field emitters. These characteristics may be very attractive for practical implementation of coherent single-electron guns. Moreover, the total energy distribution maps presented here provide direct access to states probability distributions, temperature and other parameters, and therefore can be considered as a new method of single-electron spectroscopy, alternative to measurements of capacitance[24] and conductivity[25], which are used in solid-state systems. Finally, thanks to the high charging energy and nanometer size, heterostructured carbon tips can be used to probe local electrostatic potential fields with nanometer resolution, similar to the recently developed scanning quantum dot microscopy[26,27].

## Methods

### Samples fabrication

Heterostructured carbon field emitters were fabricated by a FE-assisted structural modification of diamond needle-like crystals produced by chemical vapor deposition[28]. Here we used the same UHV setup for samples preparation and FE measurements (see next section). However, we note that similar results on FE-assisted fabrication were reproduced using another UHV setup described in ref. [29] as well. First, a diamond needle was fixed on a sharpened tungsten wire, as shown in Fig. 1a, by deposition of a metal (tungsten or platinum) contact, which provided good electrical and mechanical connection[30]. Then, the FE-assisted modification of the diamond surface was performed in the UHV setup. The diamond sample was mounted on the cathode holder and placed at a distance of 0.5 mm from a positively biased mesh gate as shown in Fig. 1b. At a gate voltage of a few hundred volts, the electric field developed at the diamond apex was enough to initiate FE. After that, the voltage cycling was performed in order to gradually increase the FE current up to values of 1-10 µA (Supplementary section I). As a result of intense Joule heating induced by high current density, the surface layer of the diamond needle with a thickness of a few nanometers was transformed into a-C. Finally, the growth of a carbon nanowire was initiated on top of the a-C layer by means of a FE-assisted diffusion of surface atoms (Supplementary section III). The dimensions of the nanowire and the resistance of the junction between the nanowire and a-C layer were controlled by changing the parameters of fabrication process.

### FE measurements

The FE measurements were performed at $5 \times 10^{-8}$ Pa pressure and at room temperature using an UHV set-up described in more detail in ref.[31]. A positive DC voltage was applied to the gate electrode using a high voltage power supply (FuG HCN 35-35000) and was measured with a multimeter (Solartron 7150plus). The FE current was measured by a picoammeter (Keithley 6485) connected to the cathode electrode. Current-voltage characteristics were obtained by ramping the voltage and registering the current with a time interval of 0.2-1 s between the measurements. Total energy distributions were measured using a spectrometer (SPECS Phoibos 100) with a hemispherical analyzer. The maps of total energy distributions were obtained by measuring a series of spectra at various voltages, with a rate and integration time of about 5 s and 1 s, respectively. Figures 2-4 in the main text show normalized maps, in which the intensity of each spectrum was normalized to unity for each voltage value (see Supplementary Section II).

**Modeling**

The modeling of the current-voltage characteristics was performed using the theory of FE in the Coulomb blockade regime, which was first proposed in ref. [11]. Moreover, here we extend this theory to model the electron energy spectra, taking into account the voltage drop inside the emitter and Joule heating.

The energy diagram of the system is shown in Fig. 2d. The FE current and electron energy spectra are determined by the probabilities, $P_N$, to find the nanowire in a state with $N$ electrons at a given voltage, $V$. According to the conventional Coulomb-blockade theory, $P_N$ can be found by solving the master equation[2], which in steady state can be written using detailed balance condition and has the following form

$$e\Gamma_N^- P_N + I_N P_N = e\Gamma_N^+ P_{N-1}. \qquad (1)$$

Here the left side corresponds to the departure of an electron from the nanowire in the $N$-electron state and contains partial FE current, $I_N$, and tunneling rate from the nanowire to the substrate, $\Gamma_N^-$. Since no electrons arrive from the gate, the right side does not contain the FE term and depends only on the tunneling rate from the substrate to nanowire $\Gamma_N^+$, which describes the arrival of an electron at the nanowire in an ($N$-1)-electron state. Tunneling rates are given by[2]

$$\Gamma_N^- = \frac{\Delta E_N}{e^2 R} \frac{1}{1 - \exp(-\Delta E_N/k_B T)},$$

$$\Gamma_N^+ = \Gamma_N^- \exp(-\Delta E_N/k_B T), \qquad (2)$$

where $R$ is substrate-nanowire junction resistance, and $\Delta E_N$ is the difference in the Coulomb energies of the system for a nanowire with $N$ and $N$-1 electrons. For the equivalent circuit of the system (inset of Fig. 2a) it is given by[11]

$$\Delta E_N = E_c(N - 1/2 - C_g V/e), \quad (3)$$

where $E_c = e^2/C$ is the charging energy, $C = C_c + C_g$ is the total capacitance, $C_g$ and $C_c$ are the capacitances of the nanowire with respect to the gate and to the rest of the cathode, respectively.

The partial FE currents, $I_N$, are given by the modified Fowler-Nordheim equation, which takes into account the high curvature of the carbon nanowire[32]

$$I_N = A F_N^2 \exp(-B/F_N - C/F_N^2), \quad (4)$$

where $A$, $B$, $C$ are the parameters determined by fitting the experimental $I(V)$ curve, and $F_N$ is the field strength at the nanowire apex in the state with $N$ electrons (Supplementary Section IV). The values of $F_N$ were determined by the numerical finite-element simulation of electric field distribution in the system, as described in Supplementary Section V. An example of such a calculation is shown in Fig. 1c.

By substituting expressions (2-4) into equation (1), we obtain the system of equations, which solution gives the probabilities $P_N$. Then, the total FE current and average tunneling rates can be calculated as the sum over all states as

$$I = \sum_N I_N P_N,$$
$$\Gamma^\pm = \sum_N \Gamma_N^\pm P_N. \quad (5)$$

Similarly, the energy spectrum is calculated as a sum over all spectra corresponding to each $N$-electron state as

$$J(\varepsilon) = \sum_N j(\varepsilon - \Delta E_N) P_N, \quad (6)$$

where $j(\varepsilon)$ is the total energy distribution shifted in energy from the cathode Fermi level ($\varepsilon = 0$) by $\Delta E_N$ (Fig. 2d). The dependence $j(\varepsilon)$ is given by the FE theory[33] and takes into account the high curvature of the carbon nanowire[34] (Supplementary Section IV). By using equations (5) and (6), we fitted the experimental data and obtained the model parameters. The values of the parameters used for the simulations in Figs. 2-4 are given in the Supplementary Section V.

**Acknowledgements**

The work was supported by Russian Science Foundation project 19-72-10067.

**Author Contributions**

V.I.K. performed the experiments and theoretical calculations, V.P. assisted with electron spectroscopy measurements, Ant.S.O. and And.S.O. performed SEM and TEM characterization, V.I.K., D.L.-H. and A.N.O. co-wrote the manuscript with input from all co-authors.

**Competing Interests statement**

The authors declare no competing interests.

**Supplementary material for "Coulomb-blockade-controlled single-electron point source"**

Victor I. Kleshch[1], Vitali Porshyn[2], Anton S. Orekhov[3], Andrey S. Orekhov[3,4],
Dirk Lützenkirchen-Hecht[2], Alexander N. Obraztsov[1,5]

[1]*Department of Physics, Lomonosov Moscow State University, Moscow 119991, Russia*

[2]*Physics Department, Faculty of Mathematics and Natural Sciences, University of Wuppertal, Wuppertal 42119, Germany*

[3]*Shubnikov Institute of Crystallography of FSRC "Crystallography and Photonics" of the Russian Academy of Sciences, Moscow 119333, Russia*

[4]*Electron Microscopy for Materials Science (EMAT), Department of Physics, University of Antwerp, Antwerp B-2020, Belgium*

[5]*Department of Physics and Mathematics, University of Eastern Finland, Joensuu 80101, Finland*

## I. Field-emission-assisted fabrication of carbon nanowire structures

The growth of a carbon nanowire on the apex of a diamond needle starts with a series of voltage cycles. In order to limit the field emission (FE) current and protect the emitter from breakdown a ballast resistor in the range from 100 GΩ to 1 MΩ was used in series with the cathode. Figure S1 shows the voltage cycles obtained for a diamond needle presented in Fig. 1a in the main text. The current-voltage characteristic, $I(V)$, during the first cycles typically exhibited a hysteresis (black curve in Fig. S1) and appeared to be unstable. In subsequent cycles, we gradually reduced the ballast resistance and increased the maximum current until $I(V)$ behavior became reproducible (red curve in Fig. S1). The stabilized current-voltage characteristic is non-linear (Fig. S1b) in Fowler-Nordheim (FN) coordinates, i.e. as $Ln(I/V^2)$ vs. $1/V$, and can be well approximated by a parabolic function (see Section IV). Moreover, the prominent periodic deviations from the parabolic curve (staircase-like pattern) appeared after a conditioning in the current range of 1-10 μA (Fig. S2a). The conditioning was repeated until stabilization of the periodic behavior.

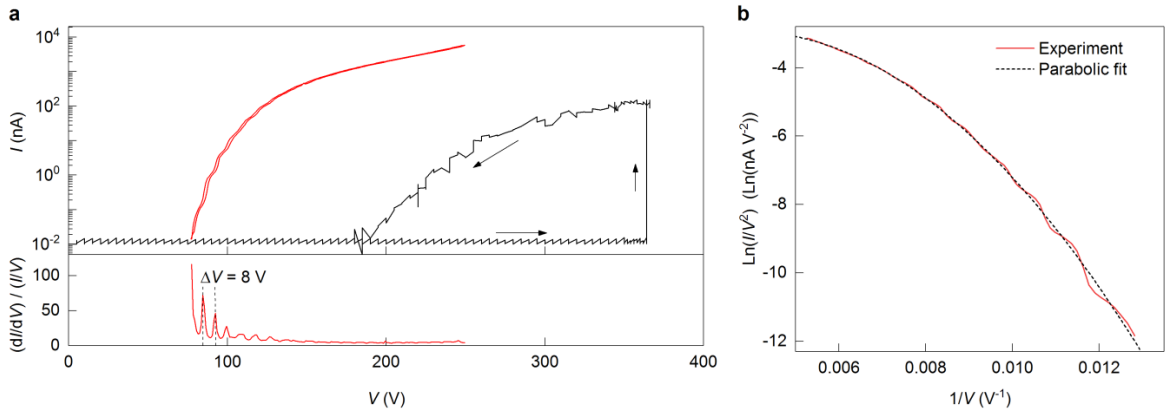

**Fig. S1. Current-voltage characteristics of a field emitter presented in Fig. 1a in the main text. (a)** Upper panel: the black curve is the first voltage cycle, obtained with a ballast resistor of 1 GΩ. The arrows represent the direction of the voltage sweeps. The red curve is the stabilized current-voltage characteristic obtained after a series of voltage cycles. Lower panel: the corresponding normalized differential conductance with a period of $\Delta V = 8$ V. **(b)** FN plot of the stabilized current-voltage characteristic from (a) and its parabolic fit.

Once a staircase $I(V)$ was obtained, we were able to change the oscillation period by exceeding a certain critical current, $I_c$, as demonstrated in Fig. S2. Figure S2a shows voltage cycle in which the staircase $I(V)$ was obtained after reaching a current of about 5 µA. This process is associated with the growth of a carbon nanowire and the formation of a heterojunction between the nanowire and the substrate (see Section III). Figure S2b shows that in the subsequent voltage cycles, when the current exceeds $I_c$~100 nA, the $I(V)$ curve may exhibit instabilities associated with changes in the geometry of the nanowire. In particular, a sharp decrease (increase) in the current can be associated with an elongation (shortening) of the nanowire. By varying the voltage in the region $I > I_c$, we were able to control the period of the resulting $I(V)$ characteristic, which was stable and reproducible after decreasing the current below $I_c$, as shown in Fig. S2c. In Fig. 2b in the main text we present the results for the same sample, which show three subsequent measurements with different periods, obtained using the described procedure. It should be noted that the critical value $I_c$ varies for different samples and depends on the growth conditions. For instance, Fig. S1a shows that the current remains stable up to ~ 7µA. At the same time, when the current level exceeded ~20 µA, the oscillation disappeared for all the samples, which is associated with the destruction of the nanowire (see Section III).

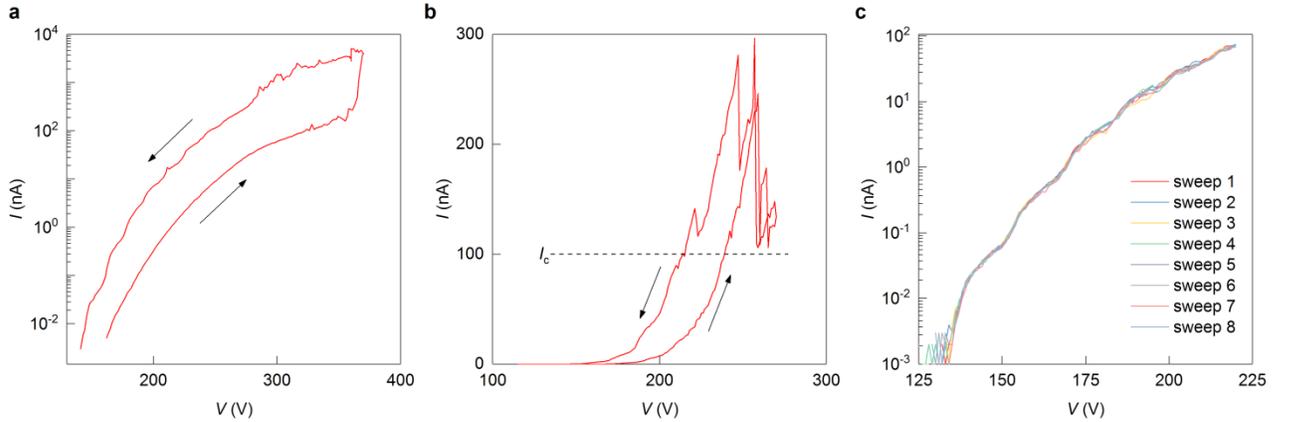

**Fig. S2. Current-voltage characteristics of the field emitter from Fig. 2 of the main text. (a)** A voltage cycle showing the appearance of the oscillations in $I(V)$ after reaching a current of about 5 µA. **(b)** $I(V)$ curve showing current instability above a critical value of $I_c$. **(c)** Reproducible $I(V)$ curves below $I_c$ obtained by eight subsequent voltage sweeps. The arrows in (a) and (b) represent the direction of the voltage sweeps.

## II. Total energy distribution maps

Total energy distributions were obtained at various applied voltages and then plotted as two-dimensional maps as shown in Fig. S3. In the main text, we present maps where each spectrum is normalized to unity, in order to clearly show the position of the maximum of each spectrum. It should be noted, that the modeling reproduces the initial spectra (before normalization), as shown in Fig. S3 b,d.

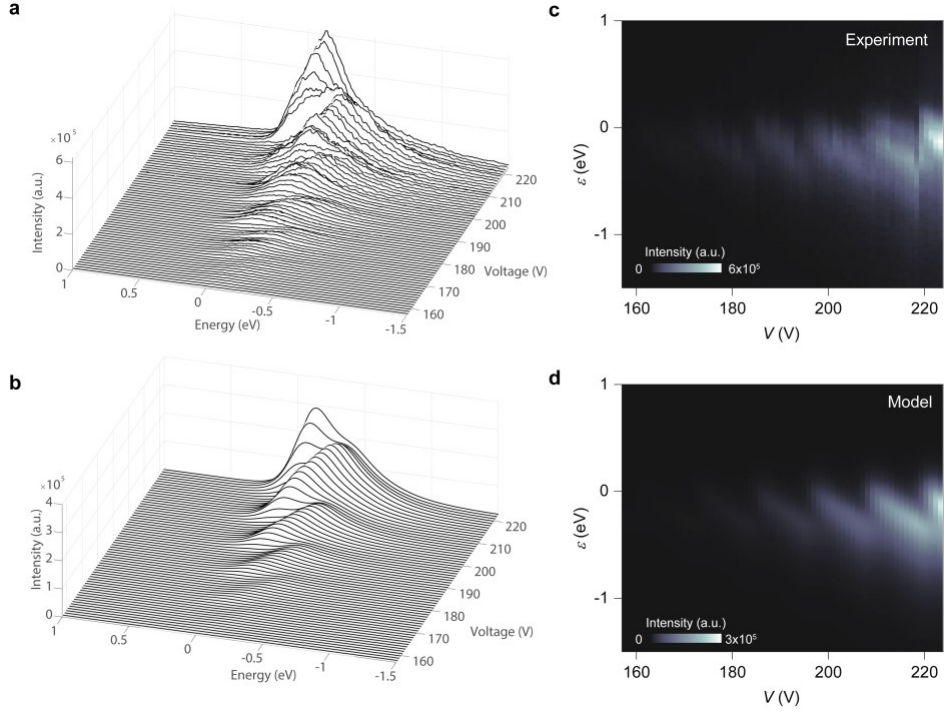

**Fig. S3. Total electron energy distributions of a field emitter from Fig. 2 of the main text. (a)** Experimental and **(b)** simulated series of total energy distributions and corresponding **(c)** experimental and **(d)** simulated 2D maps. The same maps with normalized intensities are presented in Fig. 2c and Fig. 2e in the main text.

### III. Structural analysis of carbon field emitters

Figure S4a presents a transmission electron microscopy (TEM) image of the apex region of the field emitter presented in Fig. 1a in the main text. It shows that the surface layer of a diamond needle with a thickness of about 10 nm is transformed to the amorphous carbon (a-C). Moreover, at the apex, a-C forms an elongated structure that terminates in a carbon nanowire.

The formation of microscale[1] and nanoscale[2,3] tip-shaped structures (protrusions) is a well-known phenomenon in FE from metal emitters, which is explained by surface diffusion caused by a combination of high temperature and strong electric field. In our case a similar process occurs. Firstly, the Joule heating is induced by the high current flowing through the diamond needle during a cyclic voltage change (Section I). As a result, the emitter temperature becomes high enough to allow a transformation of the surface layer from diamond to a-C. Secondly, because of the high gradient of the electric field at the apex of the ultra-sharp diamond needle, field-assisted diffusion of surface atoms occurs[2], and as a result an elongated a-C structure is formed. Finally, the carbon atoms at the apex become predominantly $sp^2$-hybridized due to the high temperature, which initiates the growth of a carbon nanowire. Such a process is possible, because it is energetically favorable for $sp^2$-carbon atoms to form tubular structures[4], rather than pyramidal structures, as in the case of metals[3]. During the growth, the field at the apex of the nanowire becomes sufficient for a field evaporation of carbon atoms[5]. Thus, the nanowire length is determined by the balance of surface diffusion and field desorption[2].

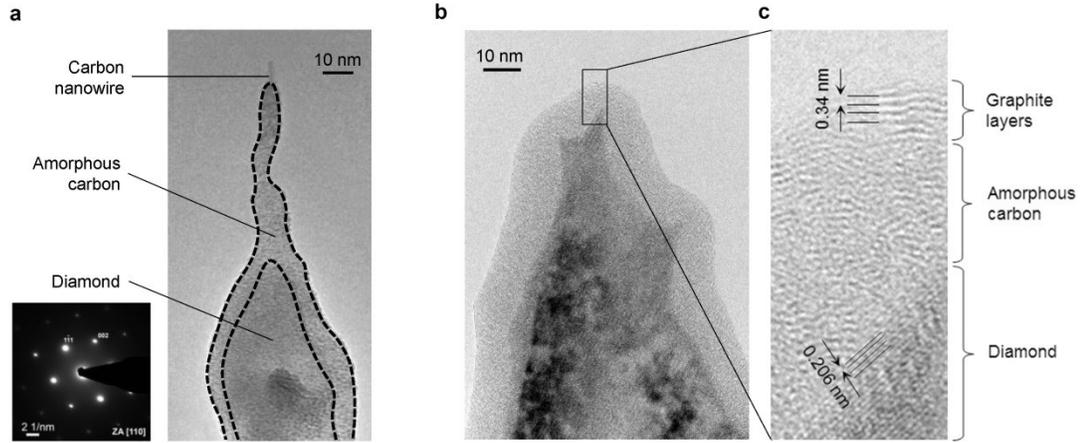

**Fig. S4. TEM analysis of carbon field emitters. (a)** TEM image of the field emitter presented in Fig. 1a in the main text. The dashed line shows the boundary of the a-C layer. The diffraction pattern shows that the structure of the inner region corresponds to a single-crystal diamond. **(b)** TEM image of the field emitter after high-current (~30 µA) processing. **(c)** High resolution TEM image of the region selected in (b). The measured interlayer distance of 0.34 nm corresponds well to (0002) plane of graphite, and an interlayer distance of 0.206 nm corresponds to (111) plane of diamond.

The diameter of the nanowire, shown in Fig. S4a and Fig. 1a, is about 1.5 nm. Its $I(V)$ characteristic, presented in Fig. S1a (red curve), shows that it can withstand a current of 7 µA. The corresponding average current density of about 4 µA/nm$^2$ is large enough to cause the Joule heating necessary to transform a-C with high sp$^3$-carbon content into graphitic layers with predominantly sp$^2$-hybridization, as demonstrated e.g. by the in situ TEM experiments[4,6]. At the same time, the current density through the relatively thick a-C layer located underneath the protrusion is much lower and, therefore, the content of sp$^3$-carbon in this region is higher. The direct imaging of the atomic structure of our nanowires was hampered by thermal and electron-beam induced vibrations. Still, it is possible to demonstrate a-C recrystallization induced by FE for emitters that were kept at currents above critical value of about 20 µA, at which the nanowires break down by overheating. A high-resolution TEM image for the sample kept at 30 µA for 10 min is shown in Fig. S4c. In this case, the surface area with ordered graphite layers can be revealed at the apex. Underneath this area an a-C layer and further the single-crystal diamond are observed. This shows that during high-current FE from the apex region, its temperature is high enough for the complete conversion of sp$^3$- to sp$^2$-carbon.

The above structural analysis shows that the content of sp$^2$-carbon in the nanowire is much higher than in the a-C substrate. Therefore, their electronic properties (work function, band gap, etc.) are substantially different. As a result, a potential barrier can be formed at the junction between the nanowire and the a-C layer. Such a possibility is demonstrated, e.g., for a-C quantum-well structures made of layers with a different sp$^2$-carbon content[7,8]. It is worth noting that we could control the resistance of the nanowire-substrate junction by choosing the maximum current level used during the FE-assisted growth. For example, for the sample with a resistance value of 3 MΩ (see Fig. 3 in the main text), the maximum current level was less than 2.5 µA, whereas the sample with $R$ = 600 kOhm (see Fig. 2 in the main text) was fabricated at about 10 µA.

## IV. Modeling of FE current and total energy distribution

In order to model the FE current, we first determined the current-voltage characteristic of the emitter, omitting the features associated with the Coulomb blockade. For that, we fitted experimental $I(V)$ curves with the modified Fowler-Nordheim (FN) formula, obtained for emitters with a small radius of curvature, and given by[9]

$$I(V) = aV^2\exp(-b/V - c/V^2). \quad (S1)$$

It differs from the standard FN formula, obtained originally for a flat surface[10], by an additional quadratic term in the exponent, which arises when the radius of the emitter becomes the same order as the width of the energy barrier at the interface with vacuum[11]. An example of the fit is shown in Fig. S1b. The parameters $a$, $b$, $c$ are obtained from the fits and used in the Coulomb-blockade model to determine the partial currents $I_N$ (see Methods in the main text).

The energy spectra were modeled using an analytical expression for the total-energy distribution of field-emitted electrons[12]

$$j(\varepsilon, F, T) = (j_{FN}(F)/d(F))e^{\varepsilon/d(F)}/(1 + e^{\varepsilon/(k_BT)}), \quad (S2)$$

where

$$j_{FN}(F) = 1.54 \times 10^{-6} \varphi^{-1} t^{-2}(y) F^2 \exp(-6.831 \times 10^{-9} \varphi^{3/2} v(y)/F) \text{ (A/m}^2\text{)}, \quad (S3)$$

$$d(F) = 9.76 \times 10^{-11} F \varphi^{-1/2} t^{-1}(y) \text{ (eV)}. \quad (S4)$$

Here, $\varepsilon$ is the kinetic electron energy relative to the Fermi level, $F$ (in V/m) is the electric field, $\varphi$ (in eV) is the work function, $j_{FN}$ is the FN equation for the current density at $T=0$ K, $v(y)$ and $t(y)$ are tabulated functions[13] of the variable $y=3.7495\times10^{-5}F^{1/2}/\varphi$. The work function was set to $\varphi = 5$ eV, which is a typical value for graphite.

The total energy distribution, $j(\varepsilon, F, T)$ at fixed $T$ and $F$ yields an asymmetric peak with a maximum near the Fermi level. The width of the spectrum increases with increasing $T$ and $F$. Equation (S.2) was obtained for a flat surface and does not take into account the small curvature of the emitter. The theory shows[11] that for a nanoscale emitter the spectrum has the same shape, while its width shrinks with a decreasing curvature radius. Fitting of the experimental spectra using equation (S.2) gives significantly underestimated values of field strength, compared to the values determined from electrostatic calculations (Section V). The values of $T$ were underestimated when compared to room temperature. Therefore, in order to simulate the FE spectra of a carbon nanowire, we replaced the temperature, $T$, and field, $F$, in Equations (S2-S4) by $k_F F$ и $k_T T$, where the coefficients $k_F$ and $k_T$ are the fitting parameters determined for each emitter. The values of the coefficients, which were used in the simulations, presented in the main text, are given in Table S1 in Section V.

## V. Modeling of electric field distribution

To model the spatial electric field distribution, we numerically solved the Laplace equation by the finite-element method in the space between the cathode and the gate. The dimensions of the model corresponded to the actual dimensions obtained using the scanning electron microscopy (SEM) images of the diamond needle and the optical images of the tungsten holder, cathode and gate. The carbon nanowire was represented by a cylinder terminated by a hemisphere.

In the case when TEM images were obtained immediately after measuring the FE characteristics, we could compare the results of the electrostatic calculations and the experiment. As an example, in Fig. S5a-c we present results of the electrostatic modeling for an emitter shown in Fig. 1a. On the basis of TEM images we built a three-dimensional (3D) model of the emitter. In the calculations, we consider the surface of the entire emitter to be equipotential at a potential of $U = 0$ V. The potential of the gate is $U = 80$ V, as in the experiment. As a result of the calculations, we obtained the potential distribution and calculated the capacitance of the nanowire with respect to the gate $C_g = 21 \pm 2$ zF. The measurement error is associated with the uncertainty in the diameter of the nanowire $D = 1.5 \pm 0.2$ nm, which was determined using TEM images. The period of Coulomb-blockade oscillations in the differential conductance for this emitter is of $\Delta V = 8$ V, as shown in Fig. S1a. The oscillation period is directly related to the nanowire-gate capacitance as $C_g = e/\Delta V$, from where one gets $C_g = 20.0$ zF in good agreement with the value calculated above.

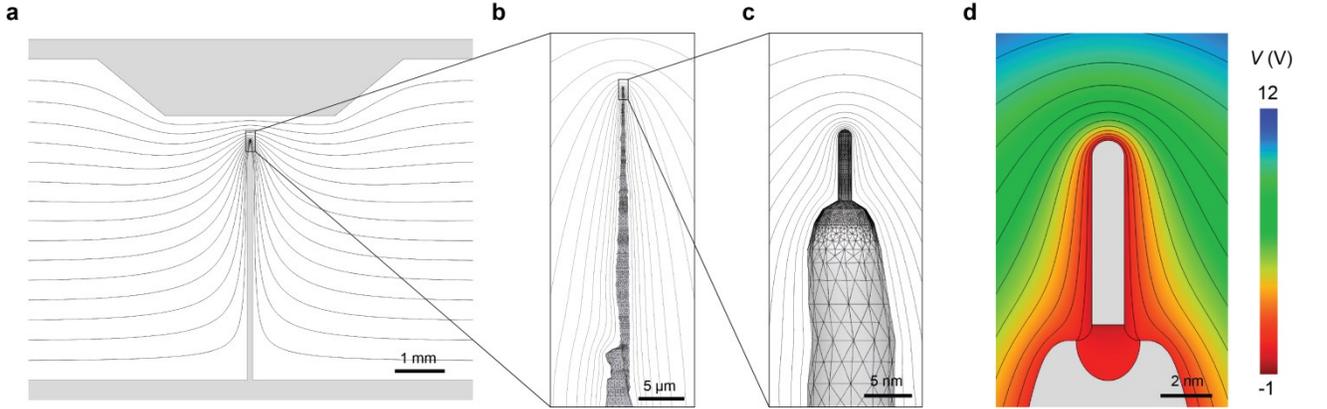

**Fig. S5. Electrostatic modeling for the field emitter presented in Fig. 1a of the main text. The gate voltage is $V = 80$ V. (a)** Cross-section of the space between the cathode and the gate. The potential difference between the equipotential lines (black) is $\Delta U = 5$ V. **(b)** 3D-mesh of a tungsten holder with a diamond needle (gray fill) and equipotential lines, $\Delta U = 3$ V. **(c)** 3D-mesh of the emitter apex and equipotential lines, $\Delta U = 1$ V. **(d)** Potential distribution for a model taking into account a depletion zone. The potential difference between the equipotential lines (black) is $\Delta U = 1$ V. The potential of the nanowire is $U = -1$ V (related to the additional charge of the nanowire), the potential of the rest of the cathode (excluding depletion zone) is $U = 0$ V, the gate potential is $U = 80$ V. The corresponding distribution of the electric field is shown in Fig. 1c in the main text.

In most experiments, the geometry of the nanowire changed during the FE experiment and could not be measured directly. In this case, the nanowire diameter, $D$, and length, $L$, were the fitting parameters. They were chosen so that the calculated values of the nanowire-gate capacitance, $C_g$, and the total capacitance, $C = C_g + C_c$ ($C_c$ – capacitance of the nanowire with respect to the rest of the cathode), coincided with the experimental values. The $C$ value is determined as $C = e^2/E_c$, where the experimental value of the charging energy, $E_c$, was obtained from the total energy distribution map. In the model, the $C_c$ value was determined as the capacitance of the Schottky barrier arising between the metal nanowire and the semiconductor a-C substrate. The possibility of Coulomb blockade in systems with a Schottky barrier, was previously demonstrated in scanning tunneling microscopy measurements, e.g. for metal nanoparticles on silicon[14]. We used the depletion zone approximation, which is valid for nanoscale Schottky contacts[15]. The depletion zone was characterized by a constant carrier density, $N_d$, and a relative permittivity, $\varepsilon_r$, and had a shape close to a hemisphere with boundary which was determined by the requirement of zero field strength outside the hemisphere[15]. It is worth noting that, due to the small size of the barrier the electron transfer via tunneling prevails over thermionic emission[15]. The relative permittivity in the surface layer of the diamond needles was previously estimated to be about $\varepsilon_r = 9$ (ref. [16]). A typical carrier concentration for a-C, $N_d = 10^{-19}$ cm$^{-3}$ (ref. [17]), was chosen for all samples, however, it is noteworthy that its variation did not significantly affect the simulation results. An example of the potential distribution calculation for a model, which takes into account the depletion zone, is presented in Fig. S5d. The corresponding distribution of the electric field is shown in Fig. 1c of the main text.

After determining the fitting parameters $L$ and $D$, we calculated the dependence of the field strength, $F$, at the apex of the nanowire on the gate voltage $V$ and the number of electrons on the nanowire, $N$, by variation of the gate potential and the charge of the nanowire. The obtained dependence $F(N, V)$ was then used to determine the FE current using equation 4 in Methods.

**Table S1. Model parameters used for simulations presented in the main text.**

| Emitter No. | Emitter I | | | Emitter II | Emitter III |
|---|---|---|---|---|---|
| Figure No. from the main text | Fig. 2a, Fig. 2b (upper panel) | Fig. 2b (middle panel) | Fig. 2b (lower panel), Fig. 2e | Fig. 3b-c | Fig. 4b-c |
| $L$ (nm) | 14.5 | 15.9 | 18.3 | 11.5 | 3.3 |
| $D$ (nm) | 1.2 | 1.2 | 1.2 | 1.5 | 1 |
| $C$ (zF) | 400.4 | 422.6 | 459.3 | 506 | 246 |
| $C_g$ (zF) | 9.47 | 10.91 | 13.47 | 7.08 | 0.62 |
| $R$ (MΩ) | 0.6 | 0.6 | 0.6 | 3 | 0.1 |
| $RC$ (ps) | 0.24 | 0.25 | 0.28 | 1.52 | 0.025 |
| $k_F$ | - | - | 4 | 2.5 | 5.9 |
| $k_T$ | - | - | 0.4 | 0.4 | 0.4 |

**VI. Fitting of the spectra and differential conductivity for emitter with small RC values**

Figure S6 shows the fitting results obtained for the emitter with $RC = 25$ fs. The fits of total energy distributions were performed as described in Section IV. For FE currents, $I$, less than 1 nA, we assumed that the emitter was at room temperature, $T = 300$ K. By fitting the spectra at $I < 1$ nA, we determined the parameter $k_T$, which was used for the fits at $I > 1$ nA. Figure S6a presents two examples of spectra at 10 nA and 100 nA, as well as their fits. The spectral broadening at 100 nA results from the increase in temperature to about 700 K. The dependence of the emitter temperature on the applied voltage is shown in Fig. S6b. The spectra with two peaks (see Fig. 4a in the main text) were excluded, because their analysis gave incorrect temperature values. By using $T(V)$ dependence we simulated $I(V)$ curves and the normalized differential conductivity.

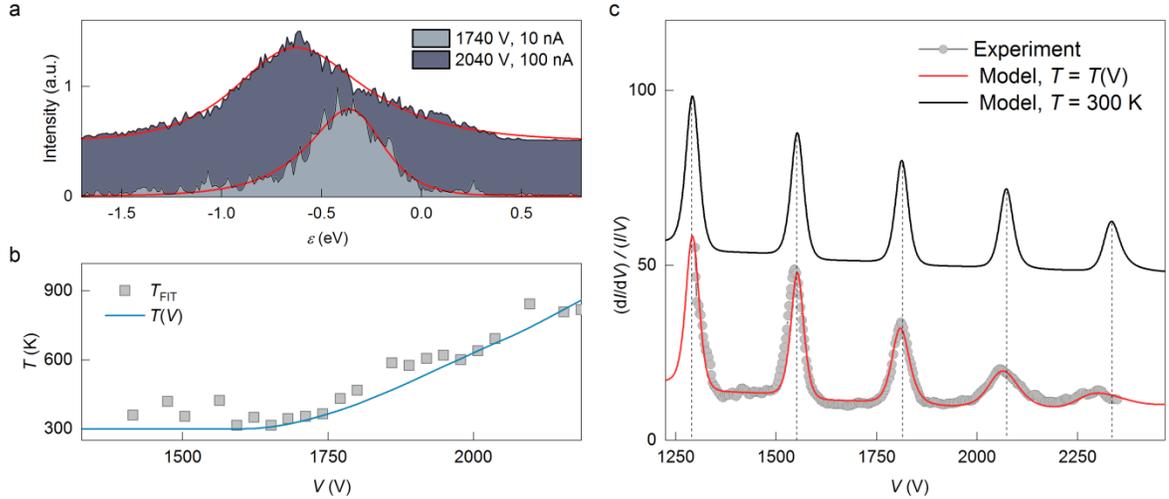

**Fig. S6. Analysis of the experimental data obtained for the emitter with $RC$=25 fs**. **(a)** Total energy distributions and corresponding fits (red curves). **(b)** Temperature values, $T_{FIT}$, obtained from fits and approximation curve, $T(V)$. **(c)** Normalized differential conductivity dependence on the applied voltage. The fit at $T = 300$ K is up-shifted by 50 for clarity. The dashed vertical lines show the positions of the peaks at $T = 300$ K.

It was found that the correct fitting of the differential conductivity oscillation, requires the introduction of an additional fractional constant charge into the model, i.e. one should replace the integer number of electrons, $N$, with $N + N_0$, where $N_0 < 1$ corresponds to the fractional charge, $N_0 e$, induced on the nanowire. The fractional residual (background) charge is a well-known parasitic phenomenon in single electron circuits[18,19]. In our case, this fractional charge can be associated with charges distributed in the depletion region[20]. It should be noted that in the case of emitters with large $RC$ values, the experimental data can be successfully fitted using $N_0$=0, since the total number of electrons $N$ in the nanowire is relatively large and the fractional $N_0$ parameter can be thus neglected.

The simulation performed at $N_0$=0.53 reproduces very accurately the experimental differential conductivity curve, as shown in Fig. S6c. It is important to note that the oscillation period is not constant and slightly decreases at high currents due to an increase in the emitter temperature. This is clearly demonstrated in Fig. 6c by the comparison with the simulation curve (black solid line) obtained using a constant temperature of 300 K.